\documentclass[
  reprint,
  aps,pra,
  superscriptaddress,
  nofootinbib,
  floatfix,
]{revtex4-2}

\usepackage{amsmath,amssymb,amsfonts,amsthm,mathtools,bm}
\usepackage{physics}
\usepackage{graphicx}
\usepackage{xcolor}
\usepackage[colorlinks,allcolors=blue]{hyperref}
\usepackage{cleveref}
\usepackage{mlmodern}

\newcommand{\bal}[1]{
\begin{align} #1 \end{align}
}

\def\lpar{\left(}
\def\rpar{\right)}
\def\lbra{\left[}
\def\rbra{\right]}

\def\circle{\scalebox{1.2}{\textbullet}}

\crefname{equation}{}{}
\Crefname{equation}{}{}
\crefmultiformat{equation}{#2(#1)#3}{–#2(#1)#3}{–#2(#1)#3}{–#2(#1)#3}
\creflabelformat{equation}{#2(#1)#3}

\newcommand{\re}[1]{\text{Re}\lpar #1 \rpar}
\newcommand{\im}[1]{\text{Im}\lpar #1 \rpar}
\renewcommand{\tr}[1]{\text{Tr}\left( #1 \right)}

\renewcommand{\braket}[1]{\left\langle #1 \right\rangle}
\newcommand{\dkl}[2]{\text{D}_{\text{KL}}\!\left( #1 \,\Vert\, #2 \right)}

\newcounter{comment}

\newcounter{example}

\newcounter{remark}
\def\cA{ {\cal A} }

\def\cC{ {\cal C} }

\def\cF{ {\cal F} }

\def\cD{ {\cal D} }

\def\cU{ {\cal U} }
\def\cX{ {\cal X} }
\def\cY{ {\cal Y} }
\def\cV{ {\cal V} }
\def\cO{ {\cal O} }
\def\cW{ {\cal W} }

\def\hc{\text{h.c.}}

\def\H{\text{H}}
\def\AH{\text{AH}}

\newcommand{\argmin}[1]{\underset{#1}{\text{argmin}}}

\def\real{\mathbb{R}}
\def\complex{\mathbb{C}}

\def\epsilon{\varepsilon}

\newcommand{\cL}{{\cal L}}
\newcommand{\be}{\begin{equation}}
\newcommand{\ee}{\end{equation}}

\newcommand{\diag}[1]{\text{diag}\left( #1 \right)}

\newtheorem{definition}{Definition}
\newtheorem{lemma}{Lemma}
\newtheorem{theorem}{Theorem}

\newtheorem{prop}{Proposition}

\begin{document}

\title{Path integral control in the complex plane}

\author{Aar\'on Villanueva}
\email{aaronv@science.ru.nl}
\affiliation{
Theoretical Division (T-4), Los Alamos National Laboratory, Los Alamos, New Mexico 87545, USA}
\affiliation{Faculty of Science, Radboud University, Heyendaalseweg 135, 6525 AJ Nijmegen, The Netherlands}

\begin{abstract}
Path integral control is a stochastic optimal control framework that computes optimal policies as weighted averages over state trajectories.
It was recently shown that, combined with stochastic quantum trajectories, the path integral control approach constitutes a promising alternative for quantum optimal control.
In this work, we propose an extension of path integral control to the complex plane using circular complex Wiener processes.
{The change of measure is defined on the driving noise space under an exact control-noise matching condition, which also accommodates rank-deficient state diffusions.}
Applied to stochastic Schr\"odinger equations, the resulting formalism gives a compact treatment of paired non-Hermitian jump channels with unequal positive dissipation rates and yields cross-entropy-optimal feedback and open-loop control formulas.
\end{abstract}

\maketitle

%================================================================================
\section{Introduction}
\label{sec:intro}

Path integral (PI) control constitutes a powerful sampling-based trajectory optimization technique for stochastic optimal control~\cite{Kappen2005,Kappen2005a,
Thijssen2015,ruiz2015adaptive}.
It enables the design of sampling schemes where the optimal controls are computed as weighted averages over sample trajectories (path integrals), thus avoiding the need to solve the Hamilton-Jacobi-Bellman equation~\cite{flemingControlledMarkovProcesses2006}.
PI control finds successful applications in robotics~\cite{raisiFaulttolerantRobustController2022,houRoboticManipulationPlanning2022,yamamotoPathIntegralPolicy2022}, autonomous driving~\cite{williamsAggressiveDrivingModel2016, gandhiRobustModelPredictive2021, mohamedAutonomousNavigationAGVs2022}, and cooperative control of multi-agent systems~\cite{broekGraphicalModelInference2008,gomezRealTimeStochasticOptimal2016,varnaiMultiAgentStochasticControl2022,wanCooperativePathIntegral2021}, among others.
Ref.~\cite{kazimRecentAdvancesPath2024} is a recent review.

More recently, PI control was combined with stochastic quantum trajectories to formulate a quantum control framework to compute open-loop policies for open and closed quantum systems~\cite{villanuevaPathIntegralControl2026}.
The framework, termed Path integral Quantum Control (PiQC), has been applied to quantum chemistry for ground state preparation and energy estimation on neutral atom simulators~\cite{najafiPathIntegralQuantum2025}.

In this Letter, we show that the PI control framework, first conceived for real-valued stochastic processes, bares a natural extension to the complex plane.
We do so by defining stochastic It\^o diffusions on the basis of complex-valued Wiener processes.
Then, building on prior work~\cite{Kappen2012, ruiz2015adaptive}, we introduce a variational framework---derived through a Kullback--Leibler (KL) formulation of the theory---{for projecting a Gibbs-tilted path measure onto a chosen control family by minimizing a cross-entropy (CE) action.}
Starting from the CE formulation, we derive CE-optimal feedback and open-loop controls and {state separately the attainability condition under which these controls also realize the unrestricted PI optimum.}

As a case study, we apply the complex framework to quantum control, thereby generalizing the specific real PiQC construction of Ref.~\cite{villanuevaPathIntegralControl2026}.
Although every complex diffusion admits an equivalent real representation, the complex notation gives a direct and structurally natural treatment of asymmetric dissipation with jump operators appearing in pairs $C_a$ and $C_a^\dagger$ and with unequal positive rates.
For deterministic open-loop controls and costs linear in the unnormalized density matrix, we establish a general mapping from the stochastic objective to deterministic control of the Lindblad equation, thereby generalizing prior work~\cite{villanuevaPathIntegralControl2026}.

The main contributions are threefold.
{First, we replace formal singular state-path densities by a Wiener-space Girsanov construction and identify exact drift matching as the essential compatibility condition.}
Second, we distinguish the forward-KL control problem from its reverse-KL CE projection and give an explicit attainability criterion.
{Third, we derive operator-level matching conditions for stochastic Schr\"odinger equations, including conjugate jump pairs with unequal positive rates.}

The text is organized as follows.
In Section~\ref{sec:complexpi} we develop the complex PI control theory.
In Sections~\ref{sec:KL_formulation} and~\ref{sec:variational} we introduce a variational formulation of the theory using KL control, and in Section~\ref{sec:feedback_control} we use it to derive feedback and open-loop controls.
In Section~\ref{sec:piqc} we apply the complex formalism to quantum optimal control and identify paired-channel models not represented by the particular real-noise construction of Ref.~\cite{villanuevaPathIntegralControl2026}.

\section{Complex path integral control}
\label{sec:complexpi}

%================================================================================

\subsection{Notation and definitions}\label{sec:definitions}

Consider the complex-valued It\^o stochastic differential equation (SDE) of the form
\bal{\label{eq:sde_general}
dX_t = f(t, X_t) dt + g_1(t, X_t) u(t, X_t) dt + g_2(t, X_t) dW_t\,.
}
Here $X_t\in\complex^N$, $f:[0,T]\times\complex^N\to\complex^N$,
$g_1:[0,T]\times\complex^N\to\complex^{N\times n_c}$, and
$g_2:[0,T]\times\complex^N\to\complex^{N\times m}$, where $n_c$ and $m$ are the numbers of control and noise channels, respectively.
The control takes values in $\complex^{n_c}$.
We regard the complex equation as an equation on $\real^{2N}$ and {assume throughout that the coefficients are Borel measurable, locally Lipschitz in the state, and of sufficient growth to give a unique nonexplosive strong solution.}
When unambiguous, we abbreviate $u_t=u(t,X_t)$.
When we refer to the function as a whole, we simply omit the time index.
The quantity $dW_t$ is an $m$-dimensional complex vector of Wiener increments of zero mean satisfying
\bal{\label{eq:noise_structure}
dW_t {dW_t}^\dagger = D_t dt\,,\quad dW_t {dW_t}^T = \Gamma_t dt\,.
}
We take $D_t > 0$ to be deterministic, positive and Hermitian.
The pseudo-covariance $\Gamma_t$ is symmetric.

Stochastic differential equations of the form \cref{eq:sde_general} represent the natural extension of real It\^o diffusions to the complex plane, where the noise structure is fully specified by $D$ and $\Gamma$.
In this work, we restrict the noise structure~\cref{eq:noise_structure} to circularly symmetric (CS) Gaussian distributions, characterized by $\Gamma \equiv 0$.

In the canonical formulation of classical path-integral control, the control acts by shifting the driving noise, so that the controlled drift
is generated in the same directions as the diffusion.
Here, the standard control-noise matching condition takes the
form of an exact factorization through a driving-noise transfer matrix.
{We use this factorization directly on Wiener space, which makes the choice of
noise-space lift explicit and remains well defined even when the
state-space diffusion is rank deficient.}

\begin{definition}[PI compatibility]\label{def:PI-compatible}
Let $\mathcal{D}\subseteq [0,T]\times\complex^N$ denote the relevant
state--time domain.  The complex diffusion~\cref{eq:sde_general} is
\emph{PI-compatible} on $\mathcal{D}$ if its driving noise is circularly
symmetric and there exists a matrix-valued map
\bal{
    G:\mathcal{D}\to\complex^{m\times n_c}\,,
}
such that $G$ satisfies
\bal{\label{eq:exact_matching}
    g_1(t,x)=g_2(t,x)G(t,x),
    \qquad (t,x)\in\mathcal{D}.
}
\end{definition}

Thus, the control $u_t$ induces the driving-noise shift
\bal{
    h_t=G(t,X_t)u_t\,,
}
and the controlled drift satisfies
\bal{
    g_1(t,X_t)u_t=g_2(t,X_t)h_t\,.
}
The usual canonical PI form, in which the dynamics are written with a
common diffusion matrix multiplying both $u_t\,dt$ and $dW_t$, is the
special case $g_1=g_2$ and $G=I$.

Define
\bal{
\label{eq:R-definitions}
    R_t &= D_t^{-1}\,,\\
    \widehat R(t,x) &= G(t,x)^\dagger R_tG(t,x)\,,\\
    dB_t &= G(t,X_t)^\dagger R_t\,dW_t\,.
}
When the underlying measure must be explicit, we write $dW_t^u$ and
$dB_t^u$ for the controlled innovations.

No rank assumption on $g_2$ is required below, because the change of
measure is formulated on the driving-noise space rather than through a
state-space density.  If $g_2$ has full column rank and
\cref{eq:exact_matching} holds, then the lift is unique and may be
written as
\bal{
    G=g_2^+g_1\,,
}
where $g_2^+$ is the Moore--Penrose pseudoinverse of $g_2$.

\begin{prop}[Realification]\label{prop:realification}
Every CS complex diffusion above has an equivalent doubled-real representation.
Writing $W=U+iV$, the covariance of $(U^T,V^T)^T$ is
\bal{\label{eq:real-covariance}
\frac12\begin{pmatrix}
\re{D}&-\im{D}\\
\im{D}&\re{D}
\end{pmatrix}dt.
}
For a complex matrix $A$, multiplication by $A$ is represented by
$\begin{psmallmatrix}\re{A}&-\im{A}\\ \im{A}&\re{A}\end{psmallmatrix}$.
\end{prop}
\begin{proof}
Expanding $D=E[(dU+i\,dV)(dU^T-i\,dV^T)]/dt$ together with
$\Gamma=E[(dU+i\,dV)(dU^T+i\,dV^T)]/dt=0$ gives~\cref{eq:real-covariance}.
The block representation follows by separating real and imaginary parts.
\end{proof}
Thus the complex formulation does not enlarge the mathematical class of real diffusions; its advantage is a compact representation that preserves the jump-operator and circular-noise structure.

Next, define the stochastic cost
\begin{multline}
\label{eq:gen-cost}
    S^u_t = \Phi(X_T) + \int_t^T \bigg( V_s\,ds + u_s^\dagger \widehat{R}_s u_s\,ds\\
        + \lpar u_s^\dagger dB_s^u + \hc \rpar \bigg)
\end{multline}
The term $\Phi(X_T) \in \real$ is the terminal cost, $V_s = V(s, X_s) \in \real$ is a running rate independent of the control, and $\widehat{R}_s \in \complex^{n_c \times n_c}$ is Hermitian positive semidefinite.
Whenever $\widehat R_s^{-1}$ is used, we additionally assume that $G_s$ has full column rank, so that $\widehat R_s$ is positive definite.
We use nondimensional costs, equivalently fixing the inverse-cost scale to one.
Thus $\Phi$, $u^\dagger\widehat Ru\,dt$, and $u^\dagger dB+\hc$ are dimensionless, while $V$ has units of inverse time.

In the interval $[t, T]$, define the expected cost
\bal{
    C[u] \coloneqq \braket{S^u_t}_u\,,
}
where the expectation $\braket{\,\cdot\,}_u$ is over trajectories generated by~\cref{eq:sde_general} with initial condition $(t,x)$.
An admissible control is a fixed predictable functional on the canonical path space for which the controlled equation is nonexplosive,
\bal{\label{eq:admissibility}
\braket{\int_t^T u_s^\dagger\widehat R_su_sds}_u<\infty,
}
the real stochastic integral in~\cref{eq:gen-cost} is a square-integrable martingale, and the positive and negative parts of the state cost are integrable.
Any likelihood introduced below must additionally satisfy the stated Novikov condition, and all weighted moments used in the CE formulas are assumed finite.
The expectation of the stochastic integral then vanishes, so
\bal{\label{eq:expected_cost}
    C[u] = \braket{\Phi(X_T) + \int_t^T \lpar V_s\,ds + u_s^\dagger \widehat{R}_s u_s\,ds \rpar}_u\,.
    }

Define the optimal value
\bal{\label{eq:opt-problem}
C_*=\inf_{u\in\cU}C[u].
}
When the infimum over the specified admissible control class $\cU$ is attained, a minimizer is denoted by $u^*$.

%================================================================================
\vspace{7pt}
\subsection{KL control formulation}
\label{sec:KL_formulation}

We formulate the KL representation directly on the driving-noise space.
This avoids introducing a density in the ambient state space, whose covariance $g_2Dg_2^\dagger$ may be singular.
Let $(\Omega,\cF,\{\cF_s\}_{s\in[t,T]},P^0)$ be a canonical filtered probability space satisfying the usual conditions and carrying a standard $m$-dimensional CS complex Brownian motion $\zeta^0$.
Define the continuous Gaussian martingale
\bal{\label{eq:inhomogeneous-noise}
W_s^0=\int_t^sD_r^{1/2}d\zeta_r^0,
}
so that $dW_s^0dW_s^{0\dagger}=D_sds$ and
$dW_s^0dW_s^{0T}=0$.
Under $P^0$, let $X$ solve the uncontrolled equation
\bal{\label{eq:uncontrolled}
    dX_s=f(s,X_s)ds+g_2(s,X_s)dW_s^0.
}
For an admissible predictable control $u$, define $h_s=G(s,X_s)u_s$ and assume the complex Novikov condition
\bal{\label{eq:novikov}
E_0\!\left[\exp\!\left(\int_t^T h_s^\dagger D_s^{-1}h_s\,ds\right)\right]<\infty.
}
The following complex Girsanov result is obtained by applying the usual real Girsanov theorem~\cite{oksendalStochasticDifferentialEquations2013} to the real and imaginary parts of $W^0$.
\begin{lemma}\label{lemm:RN}
Define
\bal{\label{eq:RN-uncontrolled}
\log\frac{dP^u}{dP^0}
&=\int_t^T\left[h_s^\dagger D_s^{-1}dW_s^0+\hc\right]\nonumber\\
&\quad-\int_t^Th_s^\dagger D_s^{-1}h_s\,ds.
}
Then $W^u$, defined by
\bal{
dW_s^u=dW_s^0-h_sds,
}
is a CS continuous Gaussian martingale with covariance $D_sds$ under $P^u$.
Moreover, exact matching~\cref{eq:exact_matching} implies that $X$ obeys the controlled dynamics~\cref{eq:sde_general} under $P^u$, and the Radon--Nikodym derivative can be written in terms of the controlled innovation as
\bal{\label{eq:RN}
    \frac{dP^u}{dP^0} = \exp\!\left\{
    \int_t^T u_s^\dagger \widehat{R}_s u_s ds
    + \int_t^T\left[u_s^\dagger dB_s^u+\hc\right]
    \right\},
}
where $dB_s^u=G(s,X_s)^\dagger D_s^{-1}dW_s^u$.
\end{lemma}

Define the state-dependent path cost
\bal{\label{eq:path-cost}
\cV_t(X)=\Phi(X_T)+\int_t^TV(s,X_s)ds
}
and the partition function
\bal{\label{eq:partition-function}
Z(t,x)\coloneqq E_0\!\left[e^{-\cV_t(X)}\right].
}
Define the Gibbs-tilted measure by
\bal{\label{eq:p*-2}
\frac{dP^*}{dP^0}=\frac{e^{-\cV_t}}{Z(t,x)}.
}
\begin{theorem}\label{th:KL_formulation}
Assume $\cV_t$ is finite $P^0$-almost surely, $0<Z(t,x)<\infty$, and all expectations below are finite.
Then every admissible control satisfying~\cref{eq:novikov} obeys
\bal{\label{eq:kl-map}
C[u] = \dkl{P^u}{P^*} - \ln Z(t, x).
}
The change-of-measure weight is
\bal{\label{eq:p*-u}
\frac{dP^*}{dP^u}=\frac{e^{-S_t^u}}{Z(t,x)},
\qquad Z(t,x)=\braket{e^{-S_t^u}}_u.
}
\end{theorem}
Proofs of Lemma~\ref{lemm:RN} and Theorem~\ref{th:KL_formulation} are given in Appendices~\ref{app:RN} and~\ref{app:KL}.

In general, the matrix $G$ depends on the state and time.
When it is deterministic, so is $\widehat R$; this permits it to be taken outside Monte Carlo averages and avoids repeatedly estimating and inverting a random weighted matrix.
We return to operator constructions with deterministic $G$ in Section~\ref{sec:non-stoch_G}.

When minimizers exist, Theorem~\ref{th:KL_formulation} gives
\bal{\label{eq:kl_equiv}
    \argmin{u \in \cU}\, C[u] \equiv \argmin{u\in \cU}\, \dkl{P^u}{P^*}.
}
More precisely,
\bal{\label{eq:C^*}
\inf_{u\in\cU}C[u]=-\ln Z(t,x)
+\inf_{u\in\cU}\dkl{P^u}{P^*}.
}
The second term need not vanish because the Gibbs measure $P^*$ need not be generated by an admissible control.
If there exists $u^*\in\cU$ such that $P^{u^*}=P^*$, then $C[u^*]=-\ln Z$ and~\cref{eq:p*-u} implies $S_t^{u^*}=-\ln Z$ almost surely, yielding the familiar zero-variance property~\cite{Thijssen2016thesis}.

\begin{prop}[Attainability]\label{prop:attainability}
Let
\bal{
L_s=E_0\!\left[\left.\frac{e^{-\cV_t(X)}}{Z(t,x)}\right|\cF_s\right]
}
and suppose its stochastic logarithm has the representation
\bal{\label{eq:optimal-noise-shift}
\frac{dL_s}{L_s}
=h_s^{*\dagger}D_s^{-1}dW_s^0+\hc
}
with the required integrability.
Then $P^*$ is attainable by $\cU$ if and only if there exists an admissible $u^*\in\cU$ satisfying
\bal{\label{eq:attainability}
G(s,X_s)u_s^*=h_s^*\,.
}
For unrestricted complex controls, pointwise invertibility of a square $G$ is sufficient.
For real controls or Markov-feedback classes, $G^{-1}h^*$ must additionally be real-valued or representable as a function of $(s,X_s)$, respectively.
\end{prop}
\begin{proof}
The density process generated by a control has stochastic logarithm
$ (Gu)^\dagger D^{-1}dW^0+\hc $.
Uniqueness of the stochastic logarithm therefore gives $P^u=P^*$ exactly when~\cref{eq:attainability} holds.
\end{proof}

%================================================================================

\vspace{7pt}
\subsection{Variational approach to KL control}
\label{sec:variational}

Solving the KL control problem~\cref{eq:kl_equiv} is not easy.
Although KL divergence is convex in an unrestricted density argument, the controlled family $\{P^u:u\in\cU\}$ need not be convex.
For the real case, Ref.~\cite{ruiz2015adaptive} used the Cross-Entropy (CE) method~\cite{deboerTutorialCrossEntropyMethod2005,botevCrossEntropyMethodOptimization2013} to construct an importance-sampling projection.
Inspired by the same technique, we derive a variational formulation for the complex case.

The CE method minimizes the reverse KL $\dkl{P^*}{P^u}$.
Define the CE projection value
\bal{\label{eq:reverse_KL_opt}
D_{\rm CE}^*=\inf_{u\in\cU}\dkl{P^*}{P^u}.
}
When this infimum is attained, a minimizer is denoted by $u_{\rm CE}$.
The forward- and reverse-KL projections generally differ for a restricted control family.
They coincide when $P^*$ is exactly attainable.

A control model $\hat{u}$ determines the class of allowed controls, which we denote by $\cU$.
For instance, given a coarse partition of $[t,T]$, the set $\cU$ can represent all piecewise-constant controls on that partition.

The variational approach to~\cref{eq:reverse_KL_opt} defines an action $\cA[\hat{u}]$ whose stationary points satisfy
\bal{
    \delta \cA = 0\,.
}
We establish the following result.
\begin{widetext}
\begin{theorem}\label{th:variational}
    Let $\cU$ be a family of complex admissible predictable controls for which the reverse KL and the cross terms below are finite, and let $\hat u_{\rm CE}$ be an interior CE minimizer.
    Up to terms independent of $\hat u$, the reverse KL equals
\bal{\label{eq:action}
    \cA[\hat{u}] \coloneqq \braket{
        \omega^u_t \int_t^T \lpar
        \hat{u}_s^\dagger \widehat{R}_s \hat{u}_s ds
        - \lbra \hat{u}_s^\dagger \widehat{R}_s u_s ds + \hat{u}_s^\dagger dB_s^u + \hc \rbra
    \rpar}_u
    }
    where $\omega^u_t=e^{-S^u_t}/Z(t,x)=dP^*/dP^u$, and $u$ is any admissible sampling control.
    For a complex variation $\delta\hat u$, the real directional derivative is twice the real part of the expression below.
    Consequently, stationarity is equivalent to
    \bal{\label{eq:variational}
    \braket{
        \omega^u_t \int_t^T
         \delta \hat{u}_s^\dagger \lpar \widehat{R}_s \hat{u}_s ds
        - \widehat{R}_s u_s ds - dB_s^u \rpar
        }_u = 0\,,
    }
    for every admissible complex variation $\delta\hat u$.
    If the weighted quadratic form induced by $\widehat R$ is strictly positive on the tangent space of $\cU$, this stationary point is the unique CE minimizer.
\end{theorem}
\end{widetext}
The proof can be found in Appendix~\ref{app:variational}.
Note that Theorem~\ref{th:variational} holds true in the continuous limit, and in the next section we use it in this sense.

%================================================================================

\vspace{7pt}
\subsection{Feedback and open-loop controls}
\label{sec:feedback_control}

The first application of the variational approach is a CE-optimal feedback projection.
A general unconstrained feedback model is given by a function $\hat{u}(s, X_s)$ satisfying~\cref{eq:variational}.
Assume $\widehat{R}$ is non-singular.
For $\Delta t>0$, define $\Delta B_{t,\Delta t}^u=\int_t^{t+\Delta t}dB_s^u$.
\begin{prop}[Initial-point CE feedback]\label{prop:fb}
    Assume the weighted increments below are integrable, localized variations may be interchanged with expectation, and the relevant coefficient averages have the required Lebesgue points.
    Then, for almost every initial time $t$, the CE-optimal feedback control at $(t,x)$ is
    \bal{\label{eq:fb}
        \hat{u}_{\rm CE}(t, x) = u(t, x) + \widehat{R}_t^{-1}
        \lim_{\Delta t\downarrow0}
        \frac{\braket{\omega_t^u\Delta B_{t,\Delta t}^u}_u}{\Delta t}
    }
    where $u$ is an arbitrary admissible sampling control.
\end{prop}

We give a proof in the Appendix~\ref{app:solving_KL}.
Brownian paths are almost surely nondifferentiable, so the increment limit in~\cref{eq:fb}, rather than a white-noise derivative, is the definition of the formula.
When $P^*$ is attainable, the CE projection coincides with the original optimum and~\cref{eq:fb} is the complex generalization of the real feedback formula in Ref.~\cite{Thijssen2015}.
The proposition is an initial-point statement.
Applying it after restarting the finite-horizon problem at each observed $(t,x)$ gives a receding-horizon feedback rule; a single global feedback function additionally requires the corresponding regular conditional construction.

Using the same techniques, it is straightforward to derive an open-loop control that is optimal in the CE sense.
Given a general unconstrained open-loop control model $\hat{u}(t)$ and an open-loop control sampler $u(t)$, we obtain the following.
\begin{prop}\label{prop:open_loop}
Assume the weighted increments are integrable, localized variations may be interchanged with expectation, and the relevant coefficient averages have the required Lebesgue points.
Then, for almost every $s$, the CE-optimal open-loop control is
\bal{\label{eq:open-loop}
    \hat{u}_{\rm CE}(s) = u(s) + \braket{\omega_0^u \widehat{R}_s}_u^{-1}
    \lim_{\Delta s\downarrow0}
    \frac{\braket{\omega_0^u \Delta B_{s,\Delta s}^u}_u}{\Delta s}\,,
}
where $\Delta B_{s,\Delta s}^u=\int_s^{s+\Delta s}dB_r^u$ and the weighted matrix is assumed invertible.
\end{prop}
The proof can be found in Appendix~\ref{app:solving_KL}.

In the next section, we apply the complex path integral control formalism to quantum control and use the variational approach to derive CE-optimal open-loop controls for state preparation.

%================================================================================

\vspace{7pt}
\section{Quantum optimal control}
\label{sec:piqc}

\subsection{PI-compatible stochastic Schr\"odinger equations}

We derive a stochastic Schr\"odinger equation (SSE) that is suitable for a path integral open-loop control formulation.
Consider the linear SSE
\begin{gather}\label{eq:linearSSEpre}
    d\psi_t = K(t)\psi_t dt + u_a(t)C_a'\psi_t dt + C_b \psi_t dW_b(t)\,,
\end{gather}
where $K$ is an operator to be specified, $C_a'\,(a=1,\ldots,n_c)$ are control operators, $C_b\,(b=1,\ldots,m)$ are jump operators, and $u_a(t)$ are open-loop controls.
Repeated indices imply summation.
We assume CS noise with $dW_t dW_t^\dagger = D_t dt$, where $D_t=\diag{\gamma_1(t),\ldots,\gamma_m(t)}$ and every $\gamma_b>0$ is deterministic.

Consider the stochastic density matrix $\rho_t = \psi_t \psi_t^\dagger$.
Average norm preservation for every initial state requires the operator identity
\bal{\label{eq:norm-condition}
K+K^\dagger+C_u'+C_u'^\dagger+\sum_b\gamma_bC_b^\dagger C_b=0,
}
where $C_u'=u_aC_a'$.
Choosing $K=-iH_0-\frac12\sum_b\gamma_bC_b^\dagger C_b$, real controls, Hermitian $H_a$, and the anti-Hermitian control drift $C_u'=-iH_u$ with $H_u=u_aH_a$ gives
\bal{\label{eq:linearSSE_final}
    d\psi_t =  -i(H_e(t) + H_u(t))\psi_t dt + C_a \psi_t dW_a(t)\,,
}
where $H_e=H_0-\frac{i}{2}\sum_a\gamma_aC_a^\dagger C_a$ is an effective non-Hermitian Hamiltonian.
Individual trajectories are not norm preserving; only their mean squared norm is conserved.

For a normalized initial state and a deterministic open-loop control, define
$\bar\rho(t)=E_u[\psi_t\psi_t^\dagger]$.
The admissibility assumptions permit expectation and time differentiation to be interchanged, and $\bar\rho$ satisfies
\bal{\label{eq:lindlbad}
    \frac{d\bar{\rho}}{dt} = \cL(\bar{\rho})\,,
}
where the Lindblad superoperator is $\cL(\bar{\rho}) = -i[H_0 + H_u, \bar{\rho}] + \cD[\cC](\bar{\rho})$, and the dissipation superoperator is $\cD[\cC](\bar{\rho}) = \sum_a \gamma_a \lpar C_a \bar{\rho} C_a^\dagger - \frac{1}{2} \{ C_a^\dagger C_a, \bar{\rho} \} \rpar$.
Thus, for deterministic open-loop controls, the linear SSE~\cref{eq:linearSSE_final} is an unnormalized unraveling of the Lindblad equation~\cite{barchielliQuantumTrajectoriesMeasurements2009}.

\begin{prop}[Lindblad objective]\label{prop:lindblad-objective}
Let $Q_T$ and $Q_s$ be deterministic positive-semidefinite operators and choose
\bal{
\Phi(\psi_T)=\braket{\psi_T|Q_T|\psi_T},
\qquad
V(s,\psi_s)=\braket{\psi_s|Q_s|\psi_s}.
}
If $u$ is deterministic and real and $\widehat R_s$ is deterministic, then the stochastic objective~\cref{eq:expected_cost} equals
\bal{\label{eq:lindblad-objective}
C[u]
&=\tr{Q_T\bar\rho(T)}\nonumber\\
&\quad+\int_0^T\left[
\tr{Q_s\bar\rho(s)}
+u_s^T\re{\widehat R_s}u_s
\right]ds.
}
\end{prop}
\begin{proof}
For a deterministic operator $Q$,
$E_u[\braket{\psi|Q|\psi}]=\tr{Q\bar\rho}$.
Apply this identity to the terminal and running terms, use deterministic $\widehat R$, and use the martingale property of the stochastic integral in~\cref{eq:gen-cost}.
\end{proof}
It gives an exact deterministic control objective for the average state.
The CE projection is nevertheless unraveling- and lift-dependent because its exponential weights depend on the full trajectory law, not only on $\bar\rho$.

For this SSE,
\bal{\label{eq:quantum-g-matrices}
g_2(\psi)&=\begin{bmatrix}C_1\psi&\cdots&C_m\psi\end{bmatrix},\nonumber\\
g_1(\psi)&=\begin{bmatrix}-iH_1\psi&\cdots&-iH_{n_c}\psi\end{bmatrix}.
}
A sufficient state-independent matching condition is the operator identity
\bal{\label{eq:operator-matching}
-iH_b=\sum_{a=1}^m C_aG_{ab},\qquad b=1,\ldots,n_c.
}
It implies $g_1(\psi)=g_2(\psi)G$ for every $\psi$, including dark states and other points where $g_2(\psi)$ loses rank.

%================================================================================
\vspace{7pt}
\subsection{Optimal pulse-based control}

In this section we derive CE-optimal open-loop controls for pulse-based quantum control.
Define a coarse partition of $[0,T]$ into $K$ subintervals $I_k=[\tau_{k-1},\tau_k)$ for $k<K$ and $I_K=[\tau_{K-1},T]$, where $\tau_0=0$ and $\tau_K=T$.
Each interval has positive length.
In a numerical discretization with step $\Delta t$, its endpoints should be chosen so that $|I_k|\geq\Delta t$.
Consider the piecewise-constant model
\bal{\label{eq:pwc}
    \hat{u}(t) = \sum_{k=1}^K \hat{A}_k \cX_k(t)
}
where $\hat{A}_k \in \real^{n_c}$ are constant pulse amplitudes, and $\cX_k$ are indicator functions of the corresponding intervals.

To derive the CE-optimal pulses $\hat{A}_{k,\rm CE}$ we extremize the action~\cref{eq:action} in the time interval $[0, T]$.
For real controls, we write the action as
\begin{multline}
    \cA[\hat{u}] =
\bigg\langle
        \omega^u_0 \int_0^T \bigg(
         \hat{u}_s^T \re{\widehat{R}_s} \hat{u}_s ds \\
        - 2\hat{u}_s^T\re{ \widehat{R}_s u_s ds + dB_s^u}
    \bigg) \bigg\rangle_u
\end{multline}
where $u$ is an arbitrary control sampler that we assume has the form~\cref{eq:pwc}, with pulse amplitudes $A_k$.
Solving the variational equation with $\hat{u}$ and $u$ piecewise constant gives
\bal{\label{eq:optimal_pulses_I}
    \hat{A}_{k,\rm CE} = A_k + \braket{\omega_0^u \re{\Delta \widehat{R}_k}}_u^{-1}
    \braket{\omega_0^u \re{\Delta B_k^u}}_u\,,
}
where $\Delta\widehat R_k=\int_{I_k}\widehat R_sds$ and $\Delta B_k^u=\int_{I_k}dB_s^u$.
The weighted matrix is assumed invertible.
Formula~\cref{eq:optimal_pulses_I} applies to real, unconstrained, piecewise-constant controls for any PI-compatible dynamics.

If $\widehat{R}$ is deterministic, the relation above reduces to
\bal{\label{eq:optimal_pulses_II}
    \hat{A}_{k,\rm CE} = A_k + \re{\Delta \widehat{R}_k}^{-1}
    \re{\braket{\omega_0^u\Delta B_k^u}_u}\,.
}
Furthermore, when $\widehat{R}$ is a constant, Eq.~\cref{eq:optimal_pulses_II} reduces to
\bal{\label{eq:optimal_pulses_III}
    \hat{A}_{k,\rm CE} = A_k + \re{\widehat{R}}^{-1}
    \re{ \frac{\braket{\omega_0^u \Delta B_k^u}_u}{\Delta \tau_k}}\,,
}
where $\Delta \tau_k = |I_k|$.
In the next section we study classes for which $G$, and therefore $\widehat R$, is deterministic.
This avoids estimating and inverting a random weighted matrix in each update.

Following~\cite{villanuevaPathIntegralControl2026}, formulas~\cref{eq:optimal_pulses_I,eq:optimal_pulses_II,eq:optimal_pulses_III} can be used in importance-sampling schemes to approximate the reverse-KL, or CE-optimal, pulses.
They minimize the original expected cost only when the Gibbs measure is attainable within the same pulse class, or when the forward- and reverse-KL projections otherwise coincide.
In the following sections, we refer to the application of complex PI control to quantum control as complex PiQC, to differentiate from real PiQC as presented in Ref.~\cite{villanuevaPathIntegralControl2026}.

%================================================================================

\vspace{7pt}
\subsection{Illustrative examples}
\label{sec:examples}

We illustrate exact matching with two single-qubit models containing excitation and relaxation channels.
These examples generalize the paired-channel construction of real PiQC in Ref.~\cite{villanuevaPathIntegralControl2026} to unequal positive rates.

To specify the objective in~\cref{eq:gen-cost}, we impose, for simplicity, a vanishing running cost, $V \equiv 0$.
The terminal cost may be any real objective satisfying the admissibility and integrability assumptions of Section~\ref{sec:definitions}; Proposition~\ref{prop:lindblad-objective} applies to objectives linear in the unnormalized density matrix.
For state preparation with a normalized target $\ket{\phi}$, define the orthogonal-complement projector
$P_\phi^\perp=I-\ket{\phi}\!\bra{\phi}$.
The nonnegative linear-unravelling objective
\bal{\label{eq:exact-fidelity-cost}
\Phi(\psi_T)=Q\braket{\psi_T|P_\phi^\perp|\psi_T},\qquad Q>0,
}
is finite almost surely and nonnegative.
Because $V=0$ in this section, the partition function defined in~\cref{eq:partition-function} is explicitly
\bal{
Z(t,x)=E_0\!\left[e^{-\Phi(\psi_T)}\right],
\qquad 0<Z(t,x)\leq1.
}
For a normalized initial state, average norm preservation gives
\bal{
\braket{\Phi(\psi_T)}_u
=Q\left[1-\tr{\ket{\phi}\!\bra{\phi}\,\bar\rho(T)}\right],
}
so minimizing the expected terminal term alone is equivalent to maximizing the Lindblad fidelity.
The complete objective also penalizes control effort and therefore trades infidelity against fluence.
A negative unnormalized quadratic fidelity is unsuitable here because its exponential moment need not be finite.
Replacing $\psi_T$ by its trajectory-wise normalization changes the objective and produces only a proxy control, which must be validated afterward through the Lindblad equation.
Likewise, observable minimization can use $\Phi(\psi_T)=\braket{\psi_T|\cO|\psi_T}$ for a positive-semidefinite observable $\cO$.

%================================================================================

\subsubsection{Excitation and relaxation with one control}
\label{sec:single_control}

Consider a qubit with $H_0=\omega\sigma_z$, one control $H_1=\sigma_x$, and the paired jump operators
\bal{
C_1=\sigma^+=(\sigma_x+i\sigma_y)/2,\qquad
C_2=\sigma^-=(\sigma_x-i\sigma_y)/2.
}
Their strictly positive rates are $\gamma_+$ and $\gamma_-$, so that
$D=\diag{\gamma_+,\gamma_-}$.
Define $Q_\gamma=\gamma_+\sigma^-\sigma^+
+\gamma_-\sigma^+\sigma^-$.
The SSE is
\begin{multline}
d\psi_t=\left[-i(\omega\sigma_z+u_t\sigma_x)
-\frac12Q_\gamma\right]\psi_tdt\\
+\sigma^+\psi_tdW_{+,t}+\sigma^-\psi_tdW_{-,t}.
\end{multline}
Because $\sigma_x=\sigma^++\sigma^-$, the operator matching condition~\cref{eq:operator-matching} holds globally with
\bal{\label{eq:G-one-control}
G_x=\begin{pmatrix}-i\\-i\end{pmatrix}.
}
Writing $r_\pm=\gamma_\pm^{-1}$ gives
\bal{
\widehat R_x&=G_x^\dagger D^{-1}G_x=r_++r_-,\nonumber\\
dB_x^u&=i(r_+dW_+^u+r_-dW_-^u).
}
Define $d\cW_s^u=r_+dW_{+,s}^u+r_-dW_{-,s}^u$.
The stochastic cost in the time interval $[0, T]$ is given by
\bal{
S_0^u=\Phi(\psi_T)+\int_0^T\left[
(r_++r_-)u_s^2ds
-2u_s\im{d\cW_s^u}
\right].
}
For the pulse model~\cref{eq:pwc}, Eq.~\cref{eq:optimal_pulses_III} becomes
\bal{
\Delta\cW_k^u=r_+\Delta W_{+,k}^u+r_-\Delta W_{-,k}^u.
}
\bal{
\hat A_{k,\rm CE}=A_k-
\frac{\im{\braket{\omega_0^u\Delta\cW_k^u}_u}}
{(r_++r_-)\Delta\tau_k}.
}
Both conjugate channels are essential.
With $C=\sigma^+$ alone and $\psi=(a,b)^T$, one has
$C\psi=(b,0)^T$ but $-i\sigma_x\psi=(-ib,-ia)^T$, so exact matching fails for generic $\psi$.
The paired model above satisfies exact drift matching for every state.
Nevertheless, $G_x:\complex\to\complex^2$ is not surjective, so a generic Gibbs noise shift $h^*$ in~\cref{eq:optimal-noise-shift} is not attainable.
The displayed pulse is therefore a CE projection unless the special attainability condition~\cref{eq:attainability} holds.

%================================================================================

\subsubsection{Relaxation/excitation with two controls}
\label{sec:two_controls}

Consider the qubit from the previous section, now equipped with two control Hamiltonians $H_1 = \sigma_x$ and $H_2 = \sigma_y$, driven by $u_x$ and $u_y$.
The jump operators correspond to the excitation and relaxation operators, $C_1 = \sigma^+$ and $C_2 = \sigma^-$, associated with constant dissipation rates, $\gamma_+$ and $\gamma_-$, respectively.
The noise matrix is given by $D = \diag{\gamma_+,\, \gamma_-}$.

The operator identities $\sigma_x=\sigma^++\sigma^-$ and
$-i\sigma_y=-\sigma^++\sigma^-$ give
\bal{\label{eq:G_single_qubit_I}
    G = \begin{pmatrix}
        -i & -1 \\
        -i & 1 \\
    \end{pmatrix}\,.
}
Because this is an operator-level factorization, it is valid for every state, including the poles of the Bloch sphere where the state-dependent diffusion matrix loses rank.
As in the one-control example, $G$ does not depend on the system state.
Writing $r=r_++r_-$ and $\Delta r=r_+-r_-$ gives
\bal{\label{eq:Rhat-two-control}
\widehat R
=\begin{pmatrix}
r&-i\Delta r\\
i\Delta r&r
\end{pmatrix},
\qquad
\re{\widehat R}=rI_2.
}
The controlled innovation is
\bal{\label{eq:dB-two-control}
dB^u
=\begin{pmatrix}
i(r_+dW_+^u+r_-dW_-^u)\\
-r_+dW_+^u+r_-dW_-^u
\end{pmatrix}.
}
Defining
$\Delta\cW_k^u=r_+\Delta W_{+,k}^u+r_-\Delta W_{-,k}^u$
and
$\Delta\cY_k^u=-r_+\Delta W_{+,k}^u+r_-\Delta W_{-,k}^u$,
the CE pulse updates are
\bal{\label{eq:two-control-updates}
\hat A_{x,k,\rm CE}
&=A_{x,k}
-\frac{\im{\braket{\omega_0^u\Delta\cW_k^u}_u}}
{r\Delta\tau_k},\nonumber\\
\hat A_{y,k,\rm CE}
&=A_{y,k}
+\frac{\re{\braket{\omega_0^u\Delta\cY_k^u}_u}}
{r\Delta\tau_k}.
}

This example works for arbitrary strictly positive rates $\gamma_+$ and $\gamma_-$.
The particular paired-channel real-noise construction of Ref.~\cite{villanuevaPathIntegralControl2026} requires equal paired rates, whereas the doubled-real representation in Proposition~\ref{prop:realification} and the present complex representation allow unequal rates.
Although $G$ is invertible over $\complex$, the Hamiltonian controls are restricted to $\real^2$.
Thus invertibility alone does not establish attainability: the Gibbs shift must lie in the real subspace $G\real^2$.

%================================================================================

\vspace{7pt}
\subsection{Operator-matched transfer matrices}
\label{sec:non-stoch_G}

The examples in Section~\ref{sec:examples} have state-independent transfer matrices because their control and jump operators obey exact linear identities.
Consider equally sized vectors of jump and control operators related by an invertible matrix $M$:
\bal{\label{eq:jump_control_relation}
    \mathbf C=M\mathbf H\,,
}
where $\mathbf C=(C_1,\ldots,C_m)^T$ and $\mathbf H=(H_1,\ldots,H_m)^T$.
If $g_H(\psi)=[H_1\psi,\ldots,H_m\psi]$, then
\bal{
g_2(\psi)=g_H(\psi)M^T,\qquad
g_1(\psi)=-ig_H(\psi).
}
It follows directly that
\bal{\label{eq:G-operator}
G=-i(M^T)^{-1}
}
satisfies $g_2(\psi)G=g_1(\psi)$ for every $\psi$.
No invertibility assumption on the state-dependent Gram matrix $J_{HH}=g_H^\dagger g_H$ is required.
At a rank-deficient state,~\cref{eq:G-operator} is a structural lift on the driving-noise space and need not equal the Moore--Penrose expression $g_2^+g_1$.

A concrete example of~\cref{eq:jump_control_relation} is the case in which the jump operators are closed under Hermitian conjugation, i.e., if $C_a$ is a jump operator, then $C_a^\dagger$ is as well.
Then the Hamiltonian controls are defined by the Hermitian and anti-Hermitian decompositions of the jump operators.
Concretely, let $C_1$ and $C_2$ be two jump operators such that $C_2 = C_1^\dagger$.
The control Hamiltonians, $H_1$ and $H_2$, are defined by the relation
\bal{\label{eq:closure_general}
    \begin{pmatrix}
        C_1 \\ C_2
    \end{pmatrix} = M \begin{pmatrix}
        H_1 \\ H_2
    \end{pmatrix}\,,
}
where
\bal{
    M = \frac{1}{2}\begin{pmatrix}
        1 & i \\
        1 & -i
    \end{pmatrix}\,.
}
We refer to this setup as Hermitian closure.
The control Hamiltonians are the full Hermitian quadratures
\bal{
H_1=C_1+C_1^\dagger,\qquad
H_2=-i(C_1-C_1^\dagger).
}
Equivalently, $H_1=2C_1^\H$ and $H_2=-2iC_1^\AH$ when the Hermitian and anti-Hermitian parts include the conventional factor $1/2$.
Note that this setup generalizes the case for the single-qubit where $C_1 = \sigma^+$ and $C_2 = \sigma^-$, where it can be verified that $G = -i (M^T)^{-1}$ and~\cref{eq:G_single_qubit_I} coincide.

The Hermitian closure setup generalizes to an arbitrary number of jump operators.
Assume $m=2p$ jump operators and $n_c=2p$ controls arranged in $p$ conjugate pairs.
Thus, each pair $C_a, C_a^\dagger$ satisfies
\bal{\label{eq:closure_pairs}
    \begin{pmatrix}
        C_a \\ C_a^\dagger
    \end{pmatrix} = M \begin{pmatrix}
        H_a^{(1)} \\ H_a^{(2)}
    \end{pmatrix} \quad a= 1,\ldots,\,p\,.
}
The matrix $G$ is block diagonal, $G=\bigoplus_{a=1}^pG_a$, where each $G_a$ is a copy of~\cref{eq:G_single_qubit_I}.

The requirement that every rate remain positive is essential for the present $D^{-1}$ formulation.
For fixed controls, sending one rate toward zero approximates the corresponding Lindblad dynamics.
It is not a regular limit of the control problem, however, because the associated inverse rate and entries of $\widehat R$ diverge.
If the remaining channels do not satisfy~\cref{eq:exact_matching}, the zero-rate model is not PI-compatible with the same controls.

%================================================================================

\vspace{7pt}
\subsubsection{Two-qubit extension with generalized amplitude damping}
\label{sec:two_qubit_extension}

As a structural extension, consider approximate preparation of
\bal{
\ket{\Phi^+}=\frac{\ket{00}+\ket{11}}{\sqrt2}
}
from $\ket{00}$.
Local controls and local jump channels cannot create entanglement by themselves, so we include an entangling system Hamiltonian, for example $H_0=J\sigma_1^z\sigma_2^z$ with $J\ne0$.
The controls are $H_{ai}=\sigma_i^a$ with fields $u_{ai}$ for $a=x,y$ and $i=1,2$.
The jump operators are $\sigma_i^\pm=(\sigma_i^x\pm i\sigma_i^y)/2$, with strictly positive rates $\gamma_i^\pm$ and
\bal{
D=\diag{\gamma_1^+,\gamma_1^-,\gamma_2^+,\gamma_2^-}.
}
This describes generalized amplitude damping.
Pure amplitude damping with exactly vanishing excitation rates is a singular limit outside the present $D^{-1}$ formulation.
With $\gamma_i^\pm>0$, bounded piecewise-continuous controls, and $T>0$, the bidirectional local dissipator is positivity improving, so $\bar\rho(T)\succ0$.
Hence the pure Bell state is not exactly reachable and serves only as a fidelity target.

The matching structure is the direct sum of two copies of the single-qubit construction in Section~\ref{sec:two_controls}.
Write the inverse of the noise matrix $D$ as $R = \diag{r^+_1\,,r^-_1,\,r^+_2\,,r^-_2}$, where $r^\pm_i = (\gamma^\pm_i)^{-1}\,(i=1,\, 2)$.
Thus, the cost matrix $\widehat{R}$ is
\bal{
    \widehat{R} = G^\dagger R G
    = \widehat{R}_1 \oplus \widehat{R}_2
}
where
\bal{\widehat{R}_i = 
    \begin{pmatrix}
    r_i & -i \Delta r_i \\
    i \Delta r_i & r_i
    \end{pmatrix}\,, \quad i=1,\, 2\,,
}
with $r_i = r^+_i + r^-_i$ and $\Delta r_i = r^+_i - r^-_i$.
The stochastic cost is
\bal{
S_0^u
&= Q\braket{\psi_T|
\left(I-\ket{\Phi^+}\!\bra{\Phi^+}\right)|\psi_T}\nonumber\\
&\quad+\int_0^T u_s^T\re{\widehat R}u_sds\nonumber\\
&\quad+2\int_0^T u_s^T\re{dB_s^u},
}
where $u = (u_{x1},\, u_{y1},\, u_{x2},\, u_{y2})^T$ and analogously for $dB^u$.
The average cost is
\bal{
C[u]
&=Q\left[1-\tr{\ket{\Phi^+}\!\bra{\Phi^+}\bar\rho(T)}\right]\nonumber\\
&\quad+\int_0^T u_s^T\re{\widehat R}u_sds.
}
This subsection specifies a compatible control model; a numerical state-preparation claim additionally requires explicit pulse optimization and Lindblad-level validation.

\begin{figure}[t]
\includegraphics[width=\columnwidth]{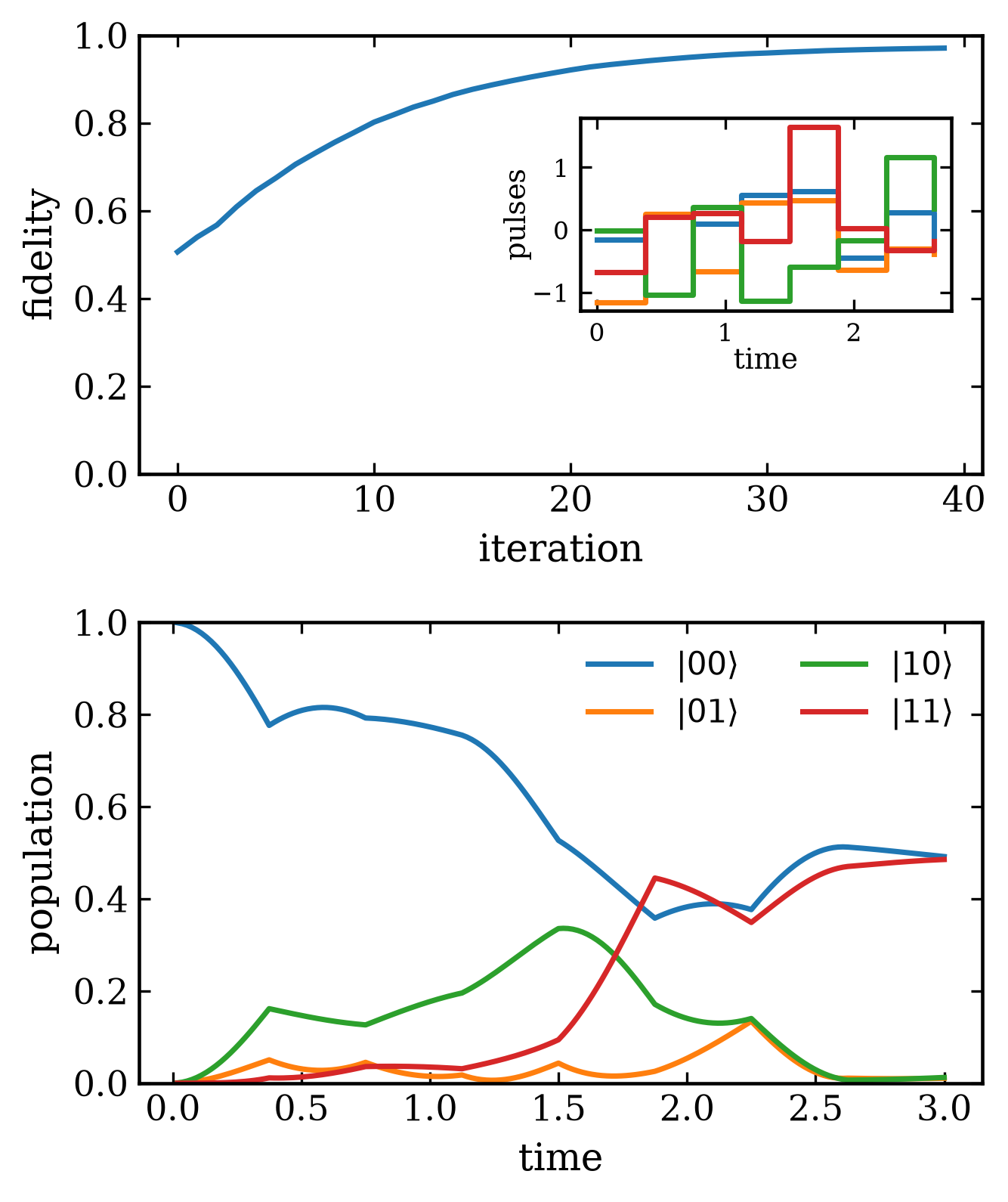}
\caption{Cross-entropy preparation of $\ket{\Phi^+}$ with generalized amplitude damping.
Top: Lindblad fidelity along the learned trajectory, with the four piecewise-constant controls inset.
Bottom: computational-basis populations of $\bar\rho(t)$.}
\label{fig:bell-summary}
\end{figure}

A numerical instance is shown in~\cref{fig:bell-summary}.
We take $J=1$, horizon $T=3$, and unequal rates $\gamma_1^+=6\times10^{-3}$, $\gamma_1^-=4\times10^{-3}$, $\gamma_2^+=7\times10^{-3}$, and $\gamma_2^-=3\times10^{-3}$.
The open-loop control is piecewise constant on eight equal subintervals of $[0,T]$.
Each iteration draws $10^3$ trajectories of the linear stochastic Schrödinger equation with Euler step $\Delta t=T/160$ and applies the constant-$\widehat R$ update~\cref{eq:optimal_pulses_III}.
The path weights are $\omega\propto\exp(-S/\lambda)$ with terminal weight $Q=15000$ and numerical temperature $\lambda=300$, chosen so that the Monte Carlo weights do not collapse onto a single trajectory.
Fidelity and the cost $C[u]$ are evaluated on the Lindblad state $\bar\rho$.
From a small random seed, forty iterations raise the terminal fidelity from the uncontrolled value $0.490$ to $0.971$.
The uncontrolled evolution of $\ket{00}$ remains separable, since $H_0$ contributes only a phase on that state and the local dissipator does not build $\ket{\Phi^+}$ on this horizon.
Under the learned pulse the populations of $\ket{00}$ and $\ket{11}$ each approach one half, while the odd-parity populations return to zero.
The notebook to reproduce this experiment can be found in~\cite{cpiqc}.

%================================================================================

\vspace{7pt}
\section{Discussion and conclusion}
\label{sec:conclusion}

We formulated a path integral control theory for complex-valued stochastic processes.
The change of measure is formulated on the driving-noise space and requires a specified exact control--noise lift $g_1=g_2G$.
This formulation remains valid when the state diffusion loses rank, while making explicit that different lifts can define different noise-space costs.
The KL identity characterizes the original problem as a forward-KL projection, whereas the variational construction yields reverse-KL, or CE-optimal, feedback and open-loop controls.
These projections coincide only under the stated attainability condition; otherwise the CE controls are unraveling-dependent importance-sampling surrogates.

Quantum control is a natural application because the theory directly accommodates complex state spaces.
Paired non-Hermitian jump channels admit an operator-level matching matrix that remains valid at dark states and allows unequal positive dissipation rates.
Proposition~\ref{prop:lindblad-objective} identifies the conditions under which the expected stochastic objective is exactly a deterministic Lindblad-control objective.
The advantage over the particular real-noise construction in Ref.~\cite{villanuevaPathIntegralControl2026} is structural rather than fundamental, as shown by the doubled-real representation in Proposition~\ref{prop:realification}.
Exactly vanishing rates, unmatched single-jump controls, performance guarantees relating the two KL projections, and numerical validation of the CE pulse updates remain outside the present scope.

This work represents a step further in the ongoing program of implementing path integral control methods for quantum optimal control~\cite{villanuevaPathIntegralControl2026,najafiPathIntegralQuantum2025}.

\section*{Acknowledgements}
This work was supported by the U.S. Department of Energy, Office of Science, Basic Energy Sciences program (award No. DE-SCL0000157).
All the contributions presented in this paper are the author's own.
The author acknowledges the use of GPT-5.6 for proofreading and assistance with code development.

\bibliography{references}
%================================================================================
\onecolumngrid

\appendix

\section{Complex Girsanov change of measure}
\label{app:RN}

Here we prove Lemma~\ref{lemm:RN}.
Set $h_s=G(s,X_s)u_s$ and consider the real continuous local martingale
\bal{
M_\tau=\int_t^\tau
\left[h_s^\dagger D_s^{-1}dW_s^0+\hc\right]
=2\re{\int_t^\tau h_s^\dagger D_s^{-1}dW_s^0}.
}
Circular symmetry gives the quadratic variation
\bal{
\langle M\rangle_\tau
=2\int_t^\tau h_s^\dagger D_s^{-1}h_sds.
}
Consequently, exponentiating the right-hand side of~\cref{eq:RN-uncontrolled} gives the stochastic exponential
$\exp\{M_\tau-\frac12\langle M\rangle_\tau\}$.
The Novikov condition~\cref{eq:novikov} makes this stochastic exponential a true martingale.
Complex Girsanov theory, equivalently real Girsanov theory applied to the real and imaginary parts, then implies that
\bal{
dW_s^u=dW_s^0-h_sds
}
is CS complex Gaussian noise with covariance $D_sds$ under $P^u$.

Substituting $dW_s^0=dW_s^u+h_sds$ in the uncontrolled dynamics~\cref{eq:uncontrolled} gives
\bal{
dX_s
&=f_sds+g_{2,s}h_sds+g_{2,s}dW_s^u\nonumber\\
&=f_sds+g_{1,s}u_sds+g_{2,s}dW_s^u,
}
where the second equality is precisely the matching condition~\cref{eq:exact_matching}.
The same substitution in~\cref{eq:RN-uncontrolled} yields
\bal{
\log\frac{dP^u}{dP^0}
&=\int_t^T\left[h_s^\dagger D_s^{-1}dW_s^u+\hc\right]
+\int_t^Th_s^\dagger D_s^{-1}h_sds\nonumber\\
&=\int_t^T\left[u_s^\dagger dB_s^u+\hc\right]
+\int_t^Tu_s^\dagger\widehat R_su_sds,
}
which proves~\cref{eq:RN}.

For comparison, a singular state-space Gaussian with covariance
$\Sigma=g_2Dg_2^\dagger$ also carries an affine-support constraint.
Put $A=g_2D^{1/2}$ and consider a matched displacement
$\delta=g_2h=A(D^{-1/2}h)$.
Its pseudoinverse quadratic form is
\bal{\label{eq:projected-state-energy}
\delta^\dagger\Sigma^+\delta
=(D^{-1/2}h)^\dagger
P_{\operatorname{Ran}(A^\dagger)}
(D^{-1/2}h)
\leq h^\dagger D^{-1}h.
}
Equality holds when $D^{-1/2}h\in\operatorname{Ran}(A^\dagger)$, in particular when $g_2$ has full column rank.
Moreover, the controlled and uncontrolled increment laws have the same affine support only when
$g_1u\in\operatorname{Ran}(g_2)$.
Thus a state-space pseudoinverse quadratic alone neither enforces exact matching nor generally reproduces the driving-noise relative entropy.

%================================================================================

\section{The KL formulation}
\label{app:KL}

We prove Theorem~\ref{th:KL_formulation}.
Combining~\cref{eq:RN,eq:p*-2} gives
\bal{
\log\frac{dP^u}{dP^*}
=\int_t^Tu_s^\dagger\widehat R_su_sds
+\int_t^T[u_s^\dagger dB_s^u+\hc]
+\cV_t(X)+\log Z(t,x).
}
The stochastic integral has zero expectation under $P^u$ by the admissibility assumptions.
Taking the $P^u$ expectation therefore gives
\bal{
\dkl{P^u}{P^*}
=C[u]+\log Z(t,x),
}
which proves~\cref{eq:kl-map}.
Finally,
\bal{
\frac{dP^*}{dP^u}
=\frac{dP^*}{dP^0}\frac{dP^0}{dP^u}
=\frac{e^{-S_t^u}}{Z(t,x)}.
}
Integrating this identity under $P^u$ proves the second equality in~\cref{eq:p*-u}.

%================================================================================

\section{The variational approach}\label{app:variational}

We prove Theorem~\ref{th:variational}.
Write the reverse KL as
\bal{
\dkl{P^*}{P^{\hat u}}
=\braket{\log\frac{dP^*}{dP^{\hat u}}}_*.
}
Terms involving $P^*$ or $P^0$ alone are independent of $\hat u$, so minimizing this expression is equivalent to minimizing
\bal{
\braket{-\log\frac{dP^{\hat u}}{dP^0}}_*.
}
For an arbitrary sampling control $u$,~\cref{eq:p*-u} changes the expectation according to
\bal{
\braket{F}_*=\braket{\omega_t^uF}_u.
}
Under $P^u$, the candidate likelihood follows from~\cref{eq:RN-uncontrolled} and $dW_s^0=dW_s^u+G_su_sds$:
\bal{
\log\frac{dP^{\hat u}}{dP^0}
=\int_t^T\left[
\hat u_s^\dagger dB_s^u
+\hat u_s^\dagger\widehat R_su_sds+\hc
\right]
-\int_t^T\hat u_s^\dagger\widehat R_s\hat u_sds.
}
It follows that the candidate-dependent part of the reverse KL is precisely
\bal{
\dkl{P^*}{P^{\hat u}}
=\cA[\hat u]+\text{constant},
}
with $\cA$ given by~\cref{eq:action}.
For a complex variation $\delta\hat u$, the real directional derivative is
\bal{\label{eq:real-directional-derivative}
D\cA[\hat u](\delta\hat u)
=2\re{\braket{
\omega_t^u\int_t^T
\delta\hat u_s^\dagger
\left[
\widehat R_s(\hat u_s-u_s)ds-dB_s^u
\right]
}_u}.
}
Equivalently, stationarity with respect to the independent Wirtinger variable $\hat u^\dagger$ gives
\bal{\label{eq:app_variational}
\braket{
        \omega^u_t \int_t^T
         \delta \hat{u}_s^\dagger \lpar \widehat{R}_s \hat{u}_s ds
        - \widehat{R}_s u_s ds - dB_s^u \rpar
        }_u = 0\,,
}
with the conjugate equation obtained by varying $\hat u$.
This proves Theorem~\ref{th:variational}.

%================================================================================

\section{Feedback and open-loop controls}
\label{app:solving_KL}

We solve the variational Eq.~\cref{eq:app_variational} for a general unconstrained feedback control model $\hat{u}(s, X_s)$.
For $\epsilon>0$, use a bounded variation localized to the interval $[t,t+\epsilon]$ and then take $\epsilon\downarrow0$.
Since $X_t=x$, the quantities $\hat u(t,x)$, $u(t,x)$, and $\widehat R(t,x)$ are non-stochastic at the initial point.
Under the weighted-integrability assumptions of Proposition~\ref{prop:fb}, dividing the stationarity equation by $\epsilon$ and applying the Lebesgue differentiation theorem gives, for almost every $t$,
\bal{
\widehat R_t\left[\hat u_{\rm CE}(t,x)-u(t,x)\right]
=\lim_{\epsilon\downarrow0}
\frac{\braket{\omega_t^u\Delta B_{t,\epsilon}^u}_u}{\epsilon},
}
where we used $\braket{\omega_t^u}_u=1$.
This proves Proposition~\ref{prop:fb}.

Now, assume we want to derive a CE-optimal open-loop control in the time interval $[0, T]$.
For a fixed $s$, use a deterministic variation supported on $[s,s+\epsilon]$.
The open-loop candidate and sampler can be taken outside the expectation.
Dividing by $\epsilon$ and applying the same theorem yields, for almost every $s$,
\bal{
\braket{\omega_0^u\widehat R_s}_u
\left[\hat u_{\rm CE}(s)-u(s)\right]
=\lim_{\epsilon\downarrow0}
\frac{\braket{\omega_0^u\Delta B_{s,\epsilon}^u}_u}{\epsilon}.
}
Inverting the weighted matrix proves Proposition~\ref{prop:open_loop}.

%================================================================================
\vspace{7pt}
\section{Exact matching and the pseudoinverse}
\label{app:G}

For the quantum dynamics,
\bal{
(g_2)_{\alpha a}=(C_a\psi)_\alpha,\qquad
(g_1)_{\alpha b}=-i(H_b\psi)_\alpha.
}
The compatibility condition is the exact linear system $g_2G=g_1$.
Multiplying by $g_2^\dagger$ gives the necessary normal equation
\bal{
J_{CC}G=J_{CH},
}
where
\bal{
(J_{CC})_{ab}=\braket{\psi|C_a^\dagger C_b|\psi},
\qquad
(J_{CH})_{ab}=-i\braket{\psi|C_a^\dagger H_b|\psi}.
}
Solving only this normal equation selects the Moore--Penrose minimum-Frobenius-norm least-squares coefficient
\bal{\label{eq:G-pseudoinverse}
G_{\rm LS}=(J_{CC})^+J_{CH}=g_2^+g_1.
}
It is a valid transfer matrix if and only if its residual vanishes:
\bal{\label{eq:matching-residual}
(I-g_2g_2^+)g_1=0.
}
Equivalently, every column of $g_1$ must lie in $\operatorname{Ran}(g_2)$.
If $g_2$ has full column rank and~\cref{eq:matching-residual} holds, the exact solution is unique and~\cref{eq:G-pseudoinverse} may be used.
At rank-deficient states an exact solution can be nonunique, and a state-dependent pseudoinverse can become unbounded near rank changes.
It is usable only when the resulting process satisfies the predictability and integrability assumptions.
The operator construction~\cref{eq:operator-matching,eq:G-operator} instead selects a state-independent structural lift.

\end{document}